\documentclass[12pt,preprint]{aastex}

\slugcomment{To appear in ApJ}

\shorttitle{Optical monitoring of QSO 2237+0305}
\shortauthors{Alcalde et al.}

\begin{document}

\title{QSO 2237+0305 $VR$ light curves from Gravitational Lenses International Time Project optical monitoring}

\author{D. Alcalde\altaffilmark{1}, E. Mediavilla\altaffilmark{1},
O. Moreau\altaffilmark{2,3}, J.\ A. Mu\~noz\altaffilmark{1},
C. Libbrecht\altaffilmark{2},
L.\ J. Goicoechea\altaffilmark{4}, J. Surdej\altaffilmark{2}, E.
Puga\altaffilmark{1,5}, Y. De Rop\altaffilmark{2}, R. Barrena\altaffilmark{1},
R. Gil--Merino\altaffilmark{4,6}, B.\ A. McLeod\altaffilmark{7},
V. Motta\altaffilmark{1,8}, A. Oscoz\altaffilmark{1} and M.
Serra--Ricart\altaffilmark{1}}

\altaffiltext{1}{Instituto de Astrof\'{\i}sica de Canarias, C/ V\'{\i}a
L\'actea s/n, E-38205 La Laguna, Tenerife, Spain; dalcalde@ll.iac.es,
emg@ll.iac.es, jmunoz@ll.iac.es,
rbarrena@ll.iac.es, vmotta@ll.iac.es, aoscoz@ll.iac.es, mserra@ot.iac.es}
\altaffiltext{2}{All\'ee du 6 Ao\^ut 17 B5c, B-4000 Sart Tilman, Belgium;
moreau@astro.ulg.ac.be, libbrech@astro.ulg.ac.be, surdej@astro.ulg.ac.be,
derop@astro.ulg.ac.be}
\altaffiltext{3}{Laboratoire d'Astronomie, Universit\'e Lille 1,
Impasse de l'Observatoire, F-59000 Lille, France}
\altaffiltext{4}{Departamento de F\'{\i}sica Moderna, Universidad de
Cantabria, Avda. de Los Castros s/n, E-39005 Santander, Cantabria, Spain;
goicol@besaya.unican.es}
\altaffiltext{5}{Present address: Max Planck Institut f\"ur Astronomie, 
K\"onigstuhl 17, Heidelberg, Germany; puga@mpia-hd.mpg.de}
\altaffiltext{6}{Lehrstuhl Astrophysik, Institut f\"ur Physik, Universit\"at 
Potsdam, Am Neuen Palais 10, D-14469 Potsdam, Germany;
rmerino@astro.physik.uni-potsdam.de}
\altaffiltext{7}{Harvard-Smithsonian Center for Astrophysics, 60 Garden
Street, Cambridge, MA 02138, USA; bmcleod@cfa.harvard.edu}
\altaffiltext{8}{Departamento de Astronom\'{\i}a, Facultad de Ciencias,
Igu\'a 4225, 11400
Montevideo, Uruguay}

\begin{abstract}
We present $VR$ observations of QSO 2237+0305 conducted by the GLITP
collaboration from 1999 October 1  to 2000 February 3. 
The observations were made with the 2.56 m Nordic Optical Telescope at Roque
de los Muchachos Observatory, La Palma (Spain). The PSF fitting 
method and an adapted version of the ISIS subtraction method have been
used to derive  the $VR$ light curves of the four components (A--D) of
the quasar. The mean errors
range in the
intervals 
0.01--0.04 mag (PSF fitting) and 0.01--0.02 mag (ISIS subtraction), with the 
faintest component (D) having the largest uncertainties. We address the 
relatively good agreement between the A-D light curves derived using
different 
filters, photometric techniques, and telescopes. The new $VR$ light curves
of component A extend the time coverage of a high magnification microlensing
peak, which was discovered by the OGLE team.
\end{abstract}

\keywords{galaxies: photometry --- gravitational lensing --- quasars:
individual (Q2237+0305)}

\section{Introduction}

Sixteen years after its discovery by Huchra and collaborators \citep{huc85}, the quadruply imaged QSO 2237+0305 is a system  that remains
 of great
theoretical and observational interest. In this system, a
high-redshift
quasar at  $z$ = 1.695 is lensed by a nearly face-on, barred spiral galaxy at
$z$ = 0.039 \citep{huc85, yee88}. The geometrical configuration of the four
lensed 
components, forming a fairly symmetrical and compact cross around the galaxy 
nucleus, implies that the light rays of the four QSO images pass through the 
bulge of the galaxy. Therefore, high optical depths to microlensing at the QSO 
image positions ($\sim 0.5$) are obtained from models \citep[e.g.,][]{sch98}.
Moreover, thanks to the unusually small distance between the observer and the
lensing galaxy,
the microlensing events should have a relatively short timescale of the order
of months. On the other hand, the time delays between the different images
derived 
from models are estimated to be of the order of hours \citep[e.g.,][]{sch98} so
that evidence in favor
of microlensing variability may be found in a direct way.

In fact, the above-mentioned theoretical expectations have been confirmed, and
the first detection of microlensing variability was made for this system 
\citep{irw89}. Other microlensing events have also been reported 
\citep[e.g.,][]{cor91, ost96, woz00a, woz00b}, and several groups are
currently
monitoring this gravitational lens system. The aim of this paper is to 
present observations from a new optical monitoring campaign obtained with the
Nordic Optical Telescope 
(NOT) within the Gravitational Lenses International Time Project (GLITP) 
collaboration. This ended monitoring program
included observations in two filters ($V$ and
$R$) with excellent temporal sampling (daily observations). Very good
seeing
conditions and angular resolutions of the cameras contributed to the obtaining
of
accurate photometry for the four lensed components (A--D) of the distant
quasar.

\section{Observations}

We observed QSO 2237+0305 from 1999 October 1  to 2000 February 3,
i.e., during approximately four
 months. All observations were made with the 2.56 m
NOT at the Roque de los Muchachos Observatory, Canary
Islands (Spain), using two different cameras. Most of the images correspond 
to the StanCam, which uses a SITe 1024$\times$1024 CCD detector with a 0.176 
arcsec/pixel scale. On some nights (at the beginning of the monitoring
period) 
we took images with the ALFOSC camera. The ALFOSC camera uses a 
Loral-Lesser 2048$\times$2048 CCD detector with a 0.188 arcsec/pixel scale. 

During good weather/technical conditions on a given night, two consecutive 300 
s exposures were taken (one in the $V$ band and another one in the $R$ 
band). The mean FWHM of the seeing disk was below 1\arcsec\ for 58\% of
nights in the $V$ filter (68\% in the $R$ filter), and the mean observing frequency in each optical band was of three
images per week.
Preprocessing of the data included the usual bias subtraction and
flat-fielding 
using dome and sky flats (when these were available). On some of the
nights,
more than one observation per filter
could be obtained. These were subsequently combined.

\section{PSF photometry}

Due to the small angular separation between the lensed components
($\approx$ 2\arcsec) and their proximity to the
galaxy nucleus, the photometry of QSO 2237+0305 is remarkably complex.
Assuming
a reasonable profile for the galaxy nucleus, one way to determine the 
brightness of the four quasar components is through PSF fitting. 

After testing exponential disks and/or de Vaucouleurs profiles and different
image sizes, we concluded that the best option for modeling the galaxy is a de 
Vaucouleurs profile within a relatively small region around the galaxy center.
Therefore, we used a model consisting of four pointlike sources and a de 
Vaucouleurs profile convolved with a PSF image, plus a constant background. 
The model was fitted to the images by adjusting its parameters to minimize the
sum of the square residuals, as described in
\citet{mcl98} and \citet{leh00}.
The model contains the following parameters: the position and intensity for
all
five sources (quasar components A--D and
galaxy), the effective radius ($R_{\rm eff}$), the 
ellipticity ($\epsilon$) and the position angle (P.A.) for the galaxy, and the
background.   

On many CCD frames, there are four field stars (those named $\alpha$ and $\beta$ in Figure 1 of 
Corrigan et al. 1991, plus two others) whose clean 2D profiles (the sky 
background has been subtracted) can be used as valid reference PSFs.
However, due to several reasons (saturation, low signal-to-noise, etc.) not all
field stars were acceptable as valid reference PSFs in all frames. We only
selected and modeled the images having  
at least two valid PSFs, i.e., 52 images (or nights) in the $V$ 
filter and 50 images in the $R$ filter. The seeing in all these images is
below 1\farcs4, so we conclude that up to this seeing level our method
works properly, as we will see in the results.

In principle, our photometric model has 19 parameters to fit. However, for
each
image, we reduced the number of free parameters to seven: quasar images
intensities, absolute position of the A component and the background. The
procedure
for fixing the other parameters was the following:
\begin{enumerate}
\item We applied the code to all images with a seeing (FWHM) better than
0\farcs7 
(7 images in the $V$ filter and 14 images in the $R$ filter) allowing all
parameters to be free. The mean values obtained for the relative positions of 
the quasar components and the galaxy are in
excellent agreement with the results obtained by the CASTLES collaboration
(Falco et al. 2002, in preparation)
using {\it Hubble Space Telescope} ({\it HST\/})---the differences in right ascension and declination are $\leq$ 
0\farcs006; see Table 1).
\item We applied the code to the images with the best seeing (FWHM $<$ 
0\farcs7), setting only the relative positions to those obtained in the 
previous step. The mean values obtained for the galaxy parameters ($R_{\rm eff}$,
$\epsilon$, and P.A.) are summarized in Table 2.
\item We applied the code to all images (whatever their seeings), setting the 
relative positions and the galaxy parameters to those derived in steps 1
and 2,
and 
allowing the remaining parameters to vary. The results 
that we obtained for the integrated galaxy intensity relative to that of
$\alpha$ star  
are distributed around a central value (6.66
in the $V$ band and 7.44 in the $R$ band) with a small dispersion (0.29
in the $V$ band and 0.26 in the $R$ band), providing further evidence
for the goodness of our fit.
\item Finally, we applied the code to all the images  now also setting the
galaxy intensity to the relative galaxy intensity (obtained in step 3) times
the $\alpha$ star intensity inferred from the code.
\end{enumerate}

Given a frame, we made several measurements of the typical instrumental flux of
each component (one for each valid reference PSF). Thus, we could use as 
the typical
instrumental flux the mean value of different PSF estimates and as error
the 
standard deviation of the mean. In this way (the PSFphotI task) we obtained one 
independent estimate of the error for each day and components  A--D. When only
two valid PSFs were available, the uncertainty in the typical instrumental
flux
of the $\alpha$ star (the calibration star; see  below) is assumed to be 
the average of the standard deviations
obtained from the days with three or four valid PSFs. However, this procedure 
is not totally consistent with the fact that the brightest reference stars
probably lead to 
better estimates of the flux than the others (the statistical weights 
being difficult to estimate). Moreover, the error measurements were obtained
from  poor statistics (typically three values for the QSO components), and
so 
the reliability of some uncertainties could not be very high. In addition, it
may be that certain 
sources of error have not been taken into account. For these reasons, we
alternatively estimated the flux of the compact sources (A--D and the $\alpha$
star) 
using only the brightest reference star that was available for PSF photometry
(the 
PSFphotII task). In this case, to compute the error in the physical flux we 
used the mean of the absolute differences between adjacent days, which is a 
very conservative estimate since it neglects the possible day-to-day
variability of the QSO components.  

To do the photometric calibration we chose the $\alpha$ star that is
always present in the images (frames), taking its $V$ magnitude from 
\citet{woz00a} and $R$ magnitude from the Guide Star Catalog II
(http://www-gsss.stsci.edu/gsc/gsc2/GSC2home.htm).

In Figure 1 we show the light curves for QSO 2237+0305 in the two bands
obtained with the variant PSFphotII. The PSFphotII task gives average errors 
of 0.01--0.04 mag in the $V$ band and 0.02--0.03 mag in the $R$ band for images 
A--D, respectively. Comparison of the PSFphotI and PSFphotII light curves
revealed that there were no significant deviations between both sets of
photometry.\footnote{The $VR$ light curves from the PSFphotI and PSFphotII tasks are available at http://www.iac.es/proyect/gravs$_-$lens/GLITP/om2237} 

\section{ISIS photometry}

In addition to the PSF photometry described in the previous section, we also
performed  
image-subtraction photometry of the four lensed components of QSO 2237+0305.
This technique 
has been pioneered by \citet{woz00a,woz00b},
using the ISIS method developed by \citet{ala00}. Because of the brightness of
the lens, 
observations of this particular
 multiply imaged quasar are well adapted for  optimal image subtraction
designed to efficiently remove the galaxy contribution without 
modeling
 its
photometric profile. To do so, we implemented in Li\`ege a locally adapted and
completed version (Moreau et al. 2002, in preparation) of the public ISIS
software provided by Alard\footnote{The ISIS 2.1 Package, http://www.iap.fr.}.

First of all, we rebinned all available CCD frames  for each of the $V$ and $R$
passbands so 
that they matched a common sampling pixel grid. A first-order 2D polynomial fit
to the 
positions of the field stars was used to correct for offsets and field
rotation;  all 
the CCD frames were then resampled on the same pixel grid using bicubic spline
interpolation.

At this step, stacking or subtracting images becomes possible for a strictly
astrometric 
application, but for an optimal co-addition, we must take into account the
disparities in 
seeing, sky background level, and possibly PSF shape from one image to another.
For both 
passbands, we built a reference frame by median-stacking eight selected images
with best 
seeing which were first convolved with an optimal 2D kernel in order to
match a
common PSF 
shape and refer to similar observing conditions. Following Alard, we modeled
the 
convolution kernels as linear combinations of three 2D Gaussian profiles, each
Gaussian 
being apodized by a 2D polynomial and then normalized in flux. The best-fit
kernel 
coefficients were determined using a least-squares algorithm taking into
account
the 
differences of the local sky background values. We then co-added these eight
convolved images 
for both passbands and built up  a deep and quasi-noiseless $V$ (and $R$)
reference
frame.

This reference frame was subsequently subtracted from each of the individual
CCD frames in 
order to obtain differential images where the only residuals expected to
appear
lie at the 
positions of the variable objects. Of course, the subtraction operation was
also optimized 
in the sense that the reference frame was successively convolved with an
adapted kernel in 
order the better to match best each individual frame. The optimal kernel was obtained
as above by 
least-squares fitting and an important point is that the coefficients of the
kernel model 
were themselves considered non-constant and also described by 2D
polynomials, enabling 
space-varying kernel solutions. Thanks to this method, the deflecting galaxy
totally 
disappears after the optimal subtraction and the observed residuals
essentially
correspond 
to the four variable lensed quasar images (details will be given in Moreau et
al. 2002, in preparation).

We could then perform a simultaneous 4-PSF fitting photometry of all
differential frames 
obtained by optimal subtraction in order to measure for each of the four
lensed
components 
the respective flux differences between each individual frame and the
reference
one. The 
photometry process begins with the determination of a mean PSF model in the
reference 
frame. We use a routine provided in ISIS which builds a normalized
composite PSF 
out of a few bright star profiles distributed all over the field. The PSF
model
obtained 
may now be adapted to each subtracted image by flux-normalized convolution
with
the 
respective optimal kernels successively determined during the subtraction
process. Using 
 original dedicated software, we then derived the flux differences from all
subtracted 
images by simultaneous least-squares fitting of four scaled PSF models
(with either positive or negative amplitudes), resampled from spline
interpolation in order 
to perfectly coincide with the exact positions of the four components forming
the gravitational lens system QSO 2237+0305.

As we are performing differential photometry, there has so far
 been no need to assume  
light-distribution models for the galaxy, with their attendant errors.
But
of course, 
in order to get the absolute fluxes at all observing epochs for each of the
four lensed 
components of QSO 2237+0305, we need to perform in both the $V$ and $R$
passbands direct 
photometry on the reference frame and, for this zero-point adjustment, we must
determine 
a model for the photometric profile of the lensing galaxy.
 
Absolute photometry of the four lensed components of QSO 2237+0305 was derived
for each 
passband using the so-called General program \citep{rem96,ost96}. We fitted
to the relevant 
part of the reference frame a global model of the gravitational lens made of
one PSF for 
each of the four lensed components plus a model of the galaxy profile. Using a
public-domain {\it HST} 
image, we measured very accurately the positions of three of the lensed
components and that 
of the galaxy center relatively to the fourth lensed component (component A), 
the position 
of which is a free parameter in the fit. For the lens galaxy model, we
adopted in the $V$ 
band a combination of a central PSF plus four Gaussian functions with
different widths, 
which led to faint and acceptable residuals. But in the case of the $R$ band,
the galaxy 
model turned out to be more complex and we fitted the galaxy using various
possible models 
(PSF + Gaussians, exponential disk, or de Vaucouleurs profiles). 
As we did not find any model clearly better than another, we averaged
the
photometric 
measurements derived for the four lensed components only taking into account
the best 
models (one PSF + two, three, or four Gaussians and one PSF + de Vaucouleurs
profile), which 
led to the faintest acceptable residuals. Having derived the flux of the four
lensed components in the reference frame (with an estimated accuracy of
0.2--2.1$\%$ in $V$ and 0.1--2.6$\%$ in $R$, for components A to D), we then
obtained their absolute fluxes at all observing epochs and, using as in
section 3 the $\alpha$ photometric calibration star, we finally constructed
their photometric light curves (see Figure 2).

The next step then consisted in deriving realistic estimates for the
photometric error 
bars. We simulated about one hundred observations for each
epoch. Of 
course, each simulated frame contained the four lensed components of QSO
2237+0305, the 
lensing galaxy and the stars in the field in order to mimic very precisely the
CCD frames 
obtained at the telescope. These simulations were made taking into account the
observed 
seeing and sky background, as well as the CCD read-out noise and photon noise.
Later on, 
we applied the adapted ISIS image-subtraction method to our simulated images.
We thus 
derived for each observing date the average and the variance of the flux for
each of the 
four lensed components and finally obtained from magnitude calibration
estimated 1$\sigma$ error bars. The averaged values of these error bars are 0.007, 0.017, 0.013 and 0.024 mag in the $V$
band and 0.005, 0.008, 0.007 and 0.013 in the $R$ band, for components A, B, C and D, respectively.\footnote{The $VR$
light curves from ISIS tasks are available at http://vela.astro.ulg.ac.be/themes/extragal/grav\-lens/bibdat/engl/lc$_-$2237.html} 
The ISIS light curves were constructed out of 53 measurements in the $V$ band and 51 in the $R$ band. Among the available observations, we rejected in fact all frames with a seeing larger than, or equal to, 1\farcs8 (i.e. 2 frames in the $V$ band and 4 in the $R$ one), plus the $V$ observation on December 13rd, because of a detector line problem exactly located on the image of QSO 2237+0305.

If we compare our absolute fluxes with those 
derived from the results obtained by PSF photometry (Section 3), we have a
difference of 0.6$\%$ for A \& B, 
2.0$\%$ for C and 4.0$\%$ for D in the $V$ band, and
0.3$\%$ for A \& B, 5.0$\%$ for C, and 9.5$\%$ for D in the $R$ band.
Let us recall however, that, here errors due to the modeling of the galaxy
profile only affect the zero point of the fluxes, not the 
general relative trend of the photometric light curves.

\section{Discussion} 

In Figure 3 we compare the light curves in the $V$ band of the four lensed
images of QSO 2237+0305 obtained from the GLITP
data (both PSF and ISIS photometry) and the OGLE data \citep{woz00b}.
The behavior of the new light curves (A--D) in the $V$ band 
basically agree with the trends reported by the OGLE collaboration during the
same 
monitoring interval, and this  strengthens both sets of photometry
(OGLE and our new dataset). There is a remarkable similarity among the three 
sets of light curves (GLITP/PSF, GLITP/ISIS, and OGLE), which involve
different photometric methods and/or telescopes. However, we notice that the 
GLITP data extend the time coverage by more than a month during a decisive
period in which a high-magnification microlensing event is observed for the A 
component. In Figures 1 and 2, a general agreement between 
the $V$ and $R$ trends can be also seen, although a full discussion on the
observed $V - R$ color gradients is beyond the scope of this introductory paper.

The two different data processing techniques (PSF and ISIS) generate error bars with different
sizes. However this fact does not imply that one approach is better than the
other one, because the estimate of the errors also depends on the method. In fact 
if we calculate the differences between consecutive nights we will find that the 
ISIS photometry has smaller dispersion than the PSF method for the A component 
but greater for the B, C, and D components.

If we examine Figure 3 in detail, we see some slight discrepancies for
several days. The trends of components  A and B are
remarkably similar and the other two components show a less good agreement. In
the three photometry sets there is  a drop for the C component, but at the end of
the
GLITP/ISIS light curve there is a rise. For the D component, the GLITP/PSF
light curve is flatter than the others. These differences could play a role in
the interpretation of the trends. On the other hand, due to the very good
time coverage in the two optical bands (e.g., from the PSF photometry we
obtained 52 final data in the $V$ band in 120 days) and the 
achievement of light curves characterized by reliable and relatively small 
photometric errors, the new dataset constitutes an important tool 
for further studies. In particular, several microlensing
analyses are now in progress and  will
appear in forthcoming papers.

\acknowledgments

The Nordic Optical Telescope is jointly operated on the island of La
Palma  by Denmark, Finland, Iceland, Norway, and Sweden, in the Spanish
Roque de Los Muchachos Observatory of the Instituto de
Astrof\'{\i}sica de
Canarias (IAC). This work was supported by project P6/88  of the IAC, 
Universidad de Cantabria funds, DGESIC (Spain) grant PB97-0220-C02, and Spanish Department of Science and Technology grant AYA2001-1647-C02. 
Research in Li\`ege was supported in part by the Belgian Office for 
Scientific, Technical and Cultural Affairs (OSTC), by PRODEX (Gravitational 
Lens Studies with HST), by contract P4/05 "P\^ole d'Attraction 
Interuniversitaire" (OSTC, Belgium), and by the "Fonds National de la 
Recherche Scientifique" (Belgium), including a two-year
post-doc research position 
for O.M. (1999-2001). We acknowledge information provided by the Guide Star Catalog II.

\clearpage

\begin{figure}
\plotone{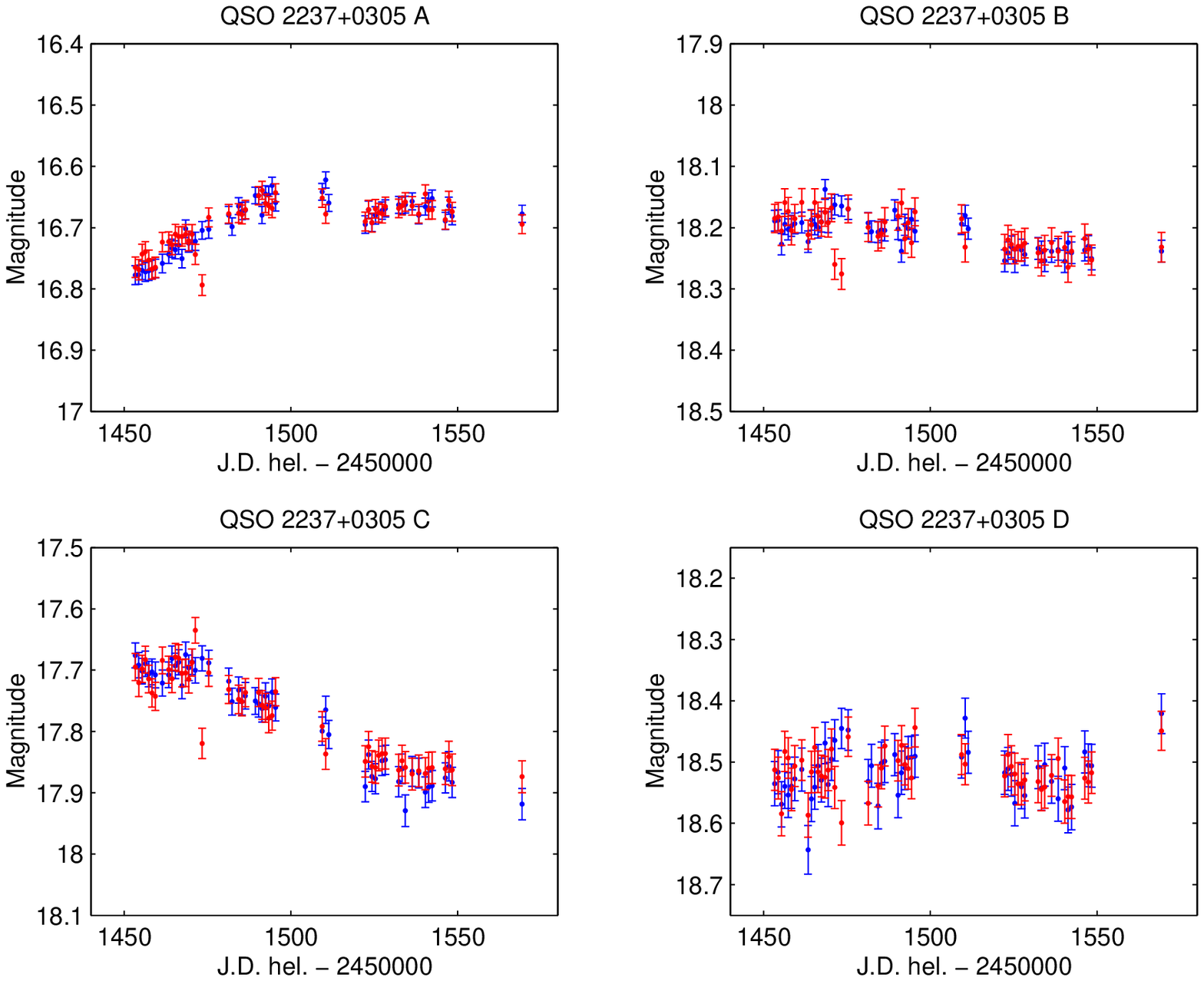}
\caption{QSO 2237+0305 light curves in the $V$ (blue dots) and $R$ (red dots)
bands obtained with  PSF photometry. We have added the mean difference
$V-R$ to the $R$ light curves for comparison.}
\end{figure}

\begin{figure}
\plotone{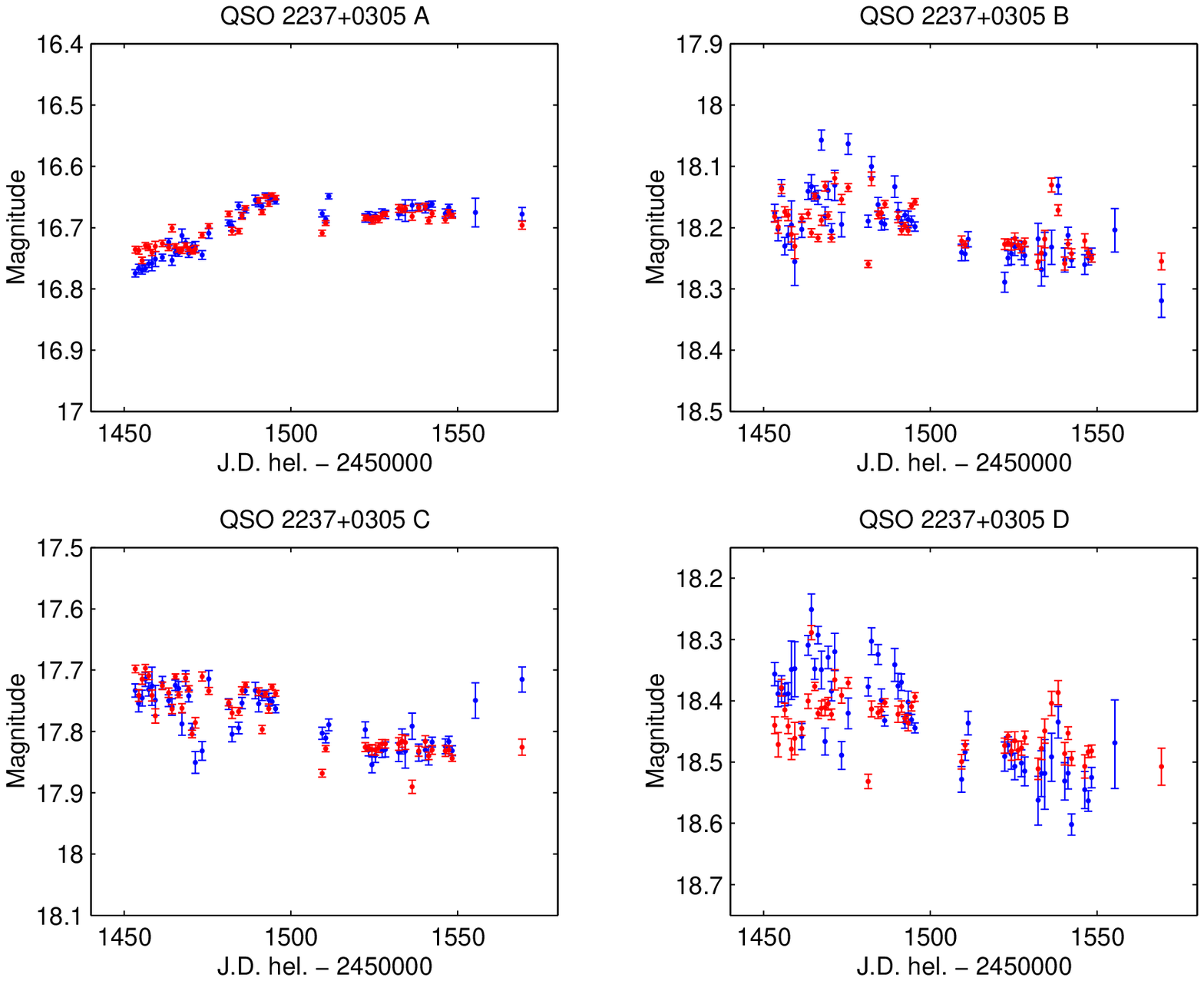}
\caption{QSO 2237+0305 light curves in the $V$ (blue dots) and $R$ (red dots)
bands obtained with  ISIS
photometry. We have added the mean difference $V-R$ to the $R$ light curves for
comparison.}
\end{figure}

\begin{figure}
\plotone{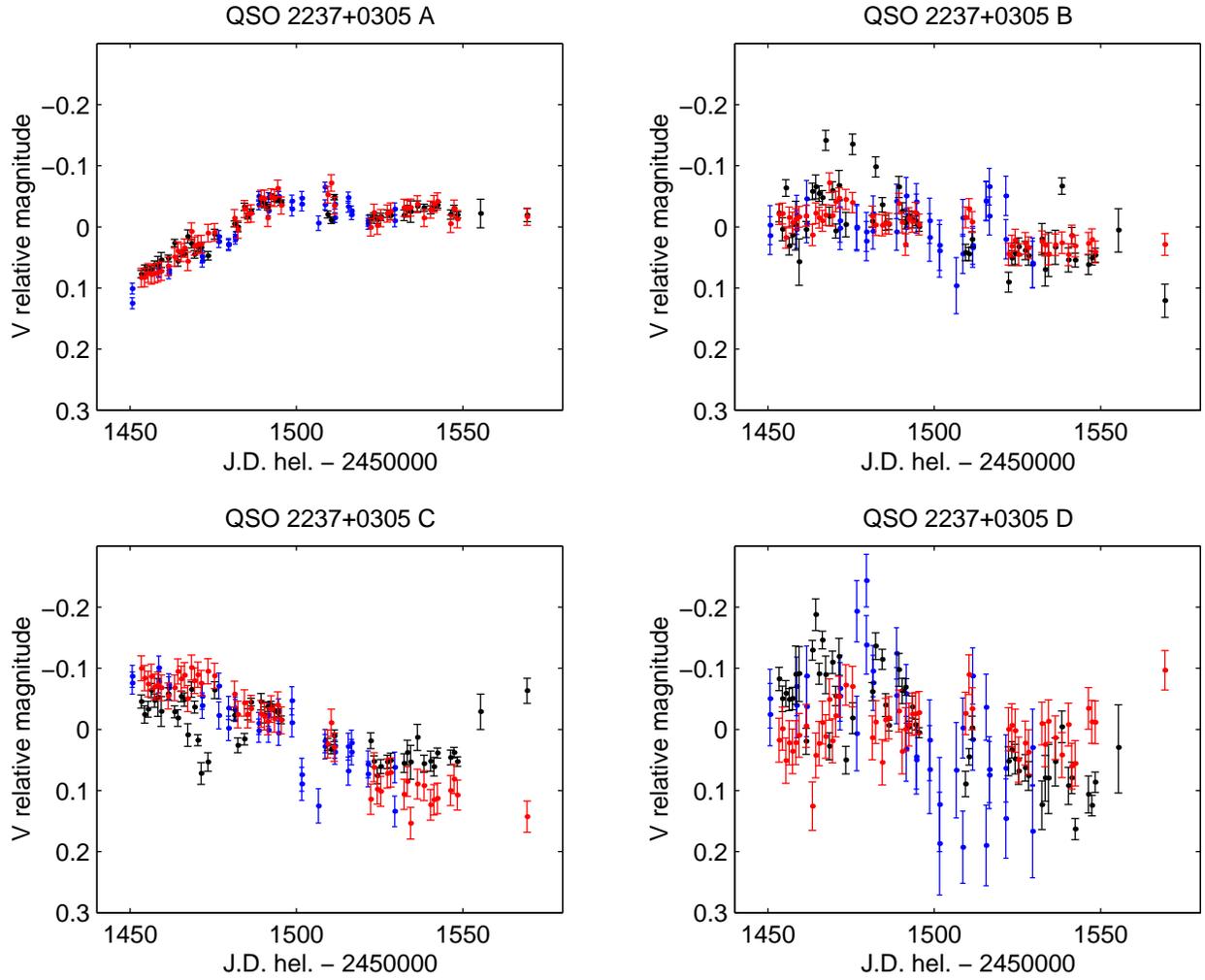}
\caption{Comparison between GLITP/PSF (red), GLITP/ISIS (black) and OGLE
(blue) QSO 2237+0305 light curves
in the $V$ band. We have subtracted the mean magnitude of each
component.\label{psfogle}}
\end{figure}

\clearpage

\begin{deluxetable}{ccccccc}
\tabletypesize{\footnotesize}
\tablecaption{Positions relative to the A component obtained by CASTLES in 
the $H$
band and by GLITP/PSF in the $R$ and $V$ bands\label{table1}}
\tablewidth{0pt}
\tablehead{\colhead{Comp.\tablenotemark{a}} &
\multicolumn{2}{c}{$H$ CASTLES}&
\multicolumn{2}{c}{$R$ GLITP/PSF}&
\multicolumn{2}{c}{$V$ GLITP/PSF}
\\
&$\Delta$R.A. (\arcsec)&$\Delta$Dec. (\arcsec)&
$\Delta$R.A. (\arcsec)&$\Delta$Dec. (\arcsec)&
$\Delta$R.A. (\arcsec)&$\Delta$Dec. (\arcsec)\\
}
\startdata
B&--0.673$\pm$0.003&1.697$\pm$0.003&--0.671$\pm$0.011&1.702$\pm$0.004&
--0.666$\pm$0.002&1.705$\pm$0.003\\
C&0.635$\pm$0.003&1.209$\pm$0.003&0.638$\pm$0.008&1.204$\pm$0.006&
0.640$\pm$0.001&1.205$\pm$0.003\\
D&--0.866$\pm$0.003&0.528$\pm$0.003&--0.868$\pm$0.007&0.531$\pm$0.005&
--0.865$\pm$0.009&0.538$\pm$0.005\\
G&--0.075$\pm$0.003&0.939$\pm$0.003&--0.075$\pm$0.010&0.942$\pm$0.010&
--0.072$\pm$0.010&0.937$\pm$0.009\\
\enddata

\tablenotetext{a}{Lensed images and galaxy.}

\end{deluxetable}

\clearpage

\begin{deluxetable}{lccc}
\tabletypesize{\normalsize}
\tablecaption{Galaxy parameters obtained by CASTLES in the $H$ band and by
GLITP/PSF
in the $R$ and $V$ bands \label{table2}}
\tablewidth{0pt}
\tablehead{\colhead{Images} &
\colhead{$R_{\rm eff}$ (\arcsec)}&
\colhead{$\epsilon$}&
\colhead{P.A. ($^{\circ}$)}}
\startdata
$H$ CASTLES&4.7$\pm$0.9&0.33$\pm$0.01&66$\pm$1\\
$R$ GLITP/PSF&4.94$\pm$0.25&0.38$\pm$0.02&62$\pm$1\\
$V$ GLITP/PSF&5.31$\pm$0.30&0.38$\pm$0.01&63$\pm$1\\
\enddata

\end{deluxetable}

\end{document}